\documentclass[sigconf]{acmart}

\ccsdesc[500]{Information systems~Recommender systems}
\ccsdesc[500]{Human-centered computing~User studies}

\copyrightyear{2025}
\acmYear{2025}
\setcopyright{None}

\begin{document}

\title{Mapping Stakeholder Needs to Multi-Sided Fairness in Candidate Recommendation for Algorithmic Hiring}
\titlenote{Accepted at the 19th ACM Conference on Recommender Systems (RecSys 2025).}

\author{Mesut Kaya}
\orcid{0000-0003-2305-6683}
\email{mesk@jobindex.dk}
\affiliation{%
  \institution{Jobindex A/S}
  \city{Copenhagen}
  \country{Denmark}
}
\affiliation{%
  \institution{IT University of Copenhagen}
  \city{Copenhagen}
  \country{Denmark}
}

\author{Toine Bogers}
\orcid{0000-0003-0716-676X}
\affiliation{%
  \institution{Department of Computer Science}
  \institution{IT University of Copenhagen}
  \city{Copenhagen} 
  \country{Denmark}
}
\email{tobo@itu.dk}

\renewcommand{\shortauthors}{Kaya and Bogers}

\begin{abstract} 
Already before the enactment of the EU AI Act, candidate or job recommendation for algorithmic hiring---semi-automatically matching CVs to job postings---was used as an example of a high-risk application where unfair treatment could result in serious harms to job seekers. 
Recommending candidates to jobs or jobs to candidates, however, is also a fitting example of a multi-stakeholder recommendation problem. In such multi-stakeholder systems, the end user is not the only party whose interests should be considered when generating recommendations. In addition to job seekers, other stakeholders---such as recruiters, organizations behind the job postings, and the recruitment agency itself---are also stakeholders in this and deserve to have their perspectives included in the design of relevant fairness metrics.
Nevertheless, past analyses of fairness in algorithmic hiring have been restricted to single-side fairness, ignoring the perspectives of the other stakeholders.
In this paper, we address this gap and present a multi-stakeholder approach to fairness in a candidate recommender system that recommends relevant candidate CVs to human recruiters in a human-in-the-loop algorithmic hiring scenario. 
We conducted semi-structured interviews with 40 different stakeholders (job seekers, companies, recruiters, and other job portal employees). We used these interviews to explore their lived experiences of unfairness in hiring, co-design definitions of fairness as well as metrics that might capture these experiences. Finally, we attempt to reconcile and map these different (and sometimes conflicting) perspectives and definitions to existing (categories of) fairness metrics that are relevant for our candidate recommendation scenario.


\end{abstract}

\keywords{Algorithmic Fairness, HR, Job Recommendation}

\maketitle

\section{Introduction}
\label{sec:intro}

The recently adopted EU AI Act\footnote{Artificial Intelligence Act (Regulation (EU) 2024/1689) \url{https://eur-lex.europa.eu/eli/reg/2024/1689}} 
represents the culmination of a growing movement towards more responsible use of AI. A core element of responsible AI is making its usage fair to individuals and groups according to some moral, legal or ethical standard by detecting and mitigating the biases that all AI-driven decision-making algorithms exhibit in order to avoid discrimination against minority groups \cite{extrand2022fairnesssurvey, barocas2023fairness}.

Person-job fit for algorithmic hiring is the process of using a computer algorithm to match candidate CVs to job postings, either for the purpose of recommending jobs to job seekers or to aid recruiters in identifying the right candidates for a job. As a `high-risk' application with the potential to impact people’s lives according to the EU AI Act, fairness in algorithmic hiring recommendation systems has also seen increased attention in recent years \cite{schumann2020we, kochling2020discriminated, raghavan2020mitigatingbias}, though with a particular focus on recommending jobs to job seekers. However, most of this work has been done at the conceptual level, where HR is often used as an example domain to illustrate fairness problems instead of designing fairness-aware algorithms \cite{extrand2022fairnesssurvey}. Actual approaches to bias detection and mitigation in algorithmic hiring are rare \cite{zehlike2017fair, geyik2019fairness, rus2022closing}, and they have focused exclusively on fairness from the job seeker’s perspective.

In reality, both job and candidate recommendation are key examples of multi-stakeholder recommendation, where the end user is not the only party whose interests are considered in generating recommendations \cite{abdollahpouri2019multi, abdollahpouri2020multistakeholder}, and where fairness concerns exist on both sides of the recommendation interaction. In the algorithmic hiring scenario at a recruitment agency,  job seekers, companies, recruiters and the recruitment agency itself are all relevant stakeholders with different needs and perspectives. Applying a conventional recommender system optimized for a single stakeholder to such a multi-stakeholder problem may result in harm to other stakeholders. In both job and candidate recommendation, unfair exposure of jobs to diverse demographic groups of job seekers or varying exposure of jobs from different types of companies could harm the benefits of job seekers, companies, recruiters, and recruitment agency.\footnote{We note here that, while fairness is important for different stakeholders, the stakes and potential harms of unfairness can vary greatly between job seekers and other stakeholders. For jobseekers, the impacts of unfairness could include discrimination based on sensitive attributes (e.g., gender or age) and the loss of or denial of job opportunities. These are severe harms. We study the importance of multi-sided fairness, without making any claim that these harms are `alike’.}

Biased or unfair decisions can appear in the design of algorithmic hiring recommender systems either due to existing biases in the data used to train these systems or biased evaluations among other reasons. In order to address the problem of fairness, identifying and measuring the bias and unfairness is a crucial step in order to mitigate biases or develop fairness-aware recommendation models when necessary. However, fairness is hard to quantify \cite{selbst2019fairness} and there is no single universally correct definition of fairness or metric to measure it  \cite{raj2022measuring}. Indeed, dozens of fairness metrics have been proposed to measure bias or (un)fairness of recommender systems \cite{raj2022measuring,smith2023scoping}. Mostly, these metrics were not designed by considering how different stakeholders may experience (un)fairness \cite{Smith2024Faact}.

As a result, there are knowledge gaps in terms of which fairness metrics are relevant for HR and how they align with the needs of the different stakeholders. We aim to take a first step towards addressing these gaps in this paper. As our use case, we consider the algorithmic hiring scenario of a recruitment agency in Denmark, where a candidate recommendation algorithm has been implemented that augments recruiters by recommending a slate of suitable candidates for an open job position. The different stakeholders involved in this recommendation scenario can have complex and often-conflicting needs and, as such, it is a compelling case study for the exploration of multi-sided fairness in algorithmic hiring.

To understand different stakeholders' understanding and perspectives of fairness in such a scenario, we conducted 40 semi-structured interviews with different stakeholders---job seekers, company representatives, recruiters and other employees of the recruitment agency---to explore their lived experiences of unfairness in hiring, co-design definitions of fairness as well as metrics that might capture these experiences. We then attempt to map their needs to fairness definitions and (categories of) metrics from the literature. We note that our goal is not to identify universally best fairness metrics for algorithmic hiring---which is an impractical task---but instead identify the metrics most suitable to our multi-stakeholder candidate recommendation scenario.

\section{Related work}
\label{sec:related-work}
\subsection{Fairness in recommendation}

The problem of how to {\em operationalize fairness} in machine learning (ML) has been studied widely in the past decade \cite{pessach2022fairness,mehrabi2021fairness,Caton2024Fairness}. Fairness operationalization consists of different stages: identifying fairness goals, defining fairness in alignment with those goals, choosing suitable fairness metrics based on these definitions, and using these metrics develop fairness-aware ML systems \cite{Jacobs2021Measuring,Stray2024Building,Smith2024Faact}. 
Fairness operationalization has also been studied extensively in field of recommender systems \cite{wang2023fairrecsys,deldjoo2024fairness,ekstrand2012fairness,raj2022measuring} and a variety of different fairness definitions and metrics have been proposed \cite{castelnovo2022clarification, Smith2024Faact, extrand2022fairnesssurvey}. A common drawback of the majority of this work is that they typically attempt to fully account for fairness only through mathematical modeling or quantitative objectives \cite{smith2023scoping}. However, fairness is a complex and multi-faceted concept that cannot be captured only quantitatively---a phenomenon also known as the {\em formalism trap}. 
This is why it is essential to also take into account {\em qualitative} fairness objectives through a more holistic understanding of these systems and their stakeholders. Participatory and collaborative methods that take into account the perspectives of all stakeholders could aid in the design of more robust and inclusive metrics, leading to a better alignment with human values through the combination of quantitative and qualitative methods \cite{jannach2020escaping, Stray2024Building, Smith2024Faact}.

\subsection{Multi-stakeholder fairness}
In recent years, there has been a growing awareness of the multi-sided or {\em multi-stakeholder} nature of ML systems and recommender systems in particular. There are many recommendation domains where other stakeholders---beyond the traditional consumers---are affected by the design and performance of the recommender system, such as music, education and HR \cite{abdollahpouri2020multistakeholder,Sonboli2021Fairness}. 
Indeed, the concept of multi-sided fairness has seen increasing attention recently \cite{abdollahpouri2019multi, abdollahpouri2020multistakeholder, sonboli2022multisided, wu2023multi, pmlr-v81-burke18a, burke2021fair, Sonboli2021Fairness,mansoury2021multisided,mansoury2021fairness, burke2025decenteringtraditionalusermultistakeholder}. 

\subsection{Fairness in algorithmic hiring}
\label{sec:fairnessHR}
One such domain where (multi-sided) fairness plays an essential role is algorithmic hiring, as evident from the literature \cite{fabris_fairness_2023, bogen2018help, kochling2020discriminated,rieskamp2023approaches,kumar2023fairness,chen2023ethics}. 
Although fairness in the HR domain has received increasing attention in recent years, there are some shortcomings of these works. So far, multi-sided fairness for the HR domain has been restricted to conceptual discussion. Any actual empirical work within multi-sided fairness in HR has focused on single stakeholders in isolation and balanced their fairness concerns with general recommendation accuracy. For instance, \citet{geyik2019fairness} proposed an algorithm to balance lists of recommended candidates based on sensitive attributes, while \citet{borisyuk2017Lijar} proposed redistributing recommended jobs to ensure fair outcomes to companies. 

Another limitation of the existing research is lack of understanding of stakeholders' perspectives of fairness, with a few exceptions. \citet{lavanchy2023applicants} examine how job seekers perceive the use of algorithms in selection and recruitment by conducting four studies on Amazon Mechanical Turk. They showed that people in the role of a job applicant perceive algorithm-driven recruitment processes as less fair compared to human only or algorithm-assisted human processes. Similarly, \citet{girona2024focusgroups} also examined job seekers' perceptions of fairness in algorithmic hiring, specifically from the perspective of minority female job seekers.

Both studies focused on a single stakeholder's perspective of fairness---the job seeker---and while they study fairness from their perspective, they do not map their findings to fairness metrics suitable for measuring fairness in the HR domain. To the best of our knowledge, we are both the first to interview different stakeholders in the algorithmic hiring domain---job seekers, recruiters, recruitment agency employees and company representatives---and the first to map their perspectives on fairness in HR to the different families of fairness metrics that are out there.

\subsection{Stakeholder perspectives on fairness}
Despite a lack of focus on multiple stakeholders, there have been several studies on the perspectives on fairness of a single stakeholder group. For example, practitioners are a commonly studied group as they are directly involved in system development \cite{holstein2019fairness,smith2023scoping,ryan2023integrating}.
For instance, \citet{holstein2019fairness} conducted 35 semi-structured interviews and anonymous survey of 267 ML practitioners to better understand industry practitioners' challenges and needs for support in developing fairer AI systems. Similarly, \citet{ryan2023integrating} interviewed 18 HCI and ML practitioners about their experiences with and perceptions of fairness. Both works aimed to understand the practitioner's perspective, but they did not map their findings to suitable fairness metrics nor guide them to select suitable metrics.  
In order to scope fairness objectives and identify suitable metrics for practitioners, \citet{smith2023scoping} iteratively designed a decision-making framework, to help practitioners identify which category of fairness metrics is most appropriate. 
\citet{Smith2024Faact} focused on collecting examples of their lived experiences with unfairness and co-designing fairness metrics with providers using focus groups.  Our study is similar to \citet{Smith2024Faact} as we also aim to map and incorporate understanding and lived experiences of stakeholders into fairness metric design, although we consider multiple different stakeholders.
%
To the best of our knowledge, the only work that aims to understand fairness considerations of different stakeholders is by \citet{smith2023multistakeholder}. They conducted semi-structured interviews with 23 employees of the Kiva micro-lending platform to understand their perspectives on fairness in micro-lending recommendation and how other stakeholder groups (lenders, borrowers and lending partners) are prioritized---though without their direct involvement.

\section{Methodology}
\label{sec:methodology}

The algorithmic hiring scenario considered in this paper is that of candidate recommendation at a Danish recruitment agency, Jobindex A/S, which owns and operates Denmark’s largest job portal and offers different levels of paid recruiting services to companies. Depending on the level, recruiters will spend more or less time searching for relevant candidates in their CV database, which contained over 160,000 CVs at the time of writing. Recruiters then contact shortlisted candidates and encourage them to apply for the job. Before the recruiters start searching the CV database, they are presented with a list of suggested candidates generated by a content-based recommender system. This system is  based on a cross-encoder architecture that takes textual input from a job posting and a CV to predict a matching score between the two. Later, these scores are used to recommend the top $N$ relevant CVs for each job posting. This algorithm was chosen to be integrated into the company's production workflow based on multiple iterations of offline and A/B testing \cite{Kaya2023}.

In this scenario, it is the recruiters who are the consumers of the recommendations and the job seekers who are the providers of the items to be recommended (i.e., their CVs) \cite{dagstuhl-burke:2024, burke2025decenteringtraditionalusermultistakeholder}. Companies are downstream stakeholders as they are impacted by the choices the recruiters make but do not directly receive recommendations. The recruitment agency itself acts as the system stakeholder, whose values may not necessarily be shared by the consumers (= recruiters), the providers (= job seekers), or the downstream stakeholders (= companies)  \cite{dagstuhl-burke:2024, burke2025decenteringtraditionalusermultistakeholder}.

Our study consisted of a total of 40 semi-structured interviews with each of these four stakeholder groups: (i) recruiters, (ii) other recruitment agency employees, (iii) job seekers, and (iv) companies. This is in line with standards for user studies in HCI \cite{Caine2016SampleSize}. 
Semi-structured interviews were the most appropriate method in our case, because they allow for detailed, rich responses on our stakeholders lived experiences with (un)fairness, on their opinions, while still offering the flexibility to brainstorm on the evaluation of fairness in algorithmic hiring.
We also considered focus groups, but in practice it turned out to be impossible to gather multiple interviewees in the same time and place. Lived experiences with unfairness can also be very personal---without the presence of a group, interviewees could feel more comfortable sharing these experiences and voicing their opinions.
In absence of an institutional review board at the recruitment agency, this research was approved by their Data Protection Officer and complied with current best research ethics practices.

\subsection{Participants}
Each interviewee is assigned an alias based on which stakeholder group they belong to. For instance, job seekers are assigned the alias JS\#, companies C\#, recruiters of the recruitment agency R\# and other employees of recruitment agency E\#. In the remainder of the paper, we will use these aliases to refer to our participants. 

\begin{table}[]
  \centering
  \footnotesize
  \begin{tabular}{cp{0.28\linewidth}cp{0.35\linewidth}}
    \toprule
    {\bf Alias} & {\bf Department} & {\bf Seniority} & {\bf Role} \\ 
    \midrule
    R1  &  Recruitment&  Senior& Recruitment manager\\ 
         R2  &  Recruitment&Senior  &Recruitment manager\\ 
         R3  & Recruitment & Senior & Team Leader \\ 
         R4 &  Recruitment&Junior  &Student researcher\\ 
         R5 &  Recruitment&Senior  &Research consultant\\ 
         R6 & Recruitment & Junior & Research consultant \\ 
         R7 & Recruitment & Junior  & Recruitment trainee \\
         R8 & Recruitment & Junior  & Recruitment consultant \\ 
         R9 & Recruitment & Junior  & Student Researcher \\ 
         R10 & Recruitment & Senior  & Recruitment consultant \\ 
    E1  & IT  &  Senior&Product manager\\ 
    E2  & Sales  &Senior  &Sales manager\\ 
    E3  &  Sales&Senior  &Sales manager\\ 
    E4  &  Employer Branding&Senior  &Copywriter\\ 
    E5  &  Sales&Senior  &Sales manager\\ 
    E6  &  Job Advertisement& Senior  &Manager\\ 
    E7 & HR & Senior  & HR manager \\ 
    E8 &Sales  &  Junior&Customer Advisor\\ 
    E9 &Sales  &Senior  &Key Account manager\\ 
    E10 &  Sales&  Junior& Customer consultant\\  
    E11 &  Sales&Senior  &Key account manager\\ 
    E12 &  HR&Senior  &HR consultant\\ 
    E13 &  Job Advertisement& Junior  &Job ad consultant\\ 
    E14 & Sales & Junior  & Job Specialist \\ 
    E15 & Employer Branding & Senior  & Employer branding consultant \\ 
    E16 & Employer Branding & Senior  & Copywriter \\ 
         \bottomrule
    \end{tabular}
    \caption{Collected demographics of recruiters and  other recruitment agency employees.}
    \label{tab:jpdemographics}
\end{table}

\subsubsection{Recruitment agency employees} 
In order to get a broad sense of challenges and needs of different departments and teams at the recruitment agency, we conducted a set of formative interviews with managers from a variety of different departments of the recruitment agency: {\em HR}, {\em Sales}, {\em Matching \& Recruitment}, {\em Job Advertisement}, and {\em Employer Branding}.
We then asked these managers and team leaders to provide us with the names of other employees that would be relevant to interview, resulting in a total of 16 interview with non-recruiter employees of the recruitment agency, the demographic characteristics of which are shown in \mbox{Table \ref{tab:jpdemographics}}.

\subsubsection{Recruiters} 
Team leaders in the {\em Matching \& Recruitment} departments provided us with 10 recruiters to interview. These recruiters varied in terms of seniority, but also which industry they specialized in. Demographic details for our recruiters are included in 
\mbox{Table \ref{tab:jpdemographics}}.

\begin{table}[]
  \centering
  \footnotesize
  \begin{tabular}{ccp{0.25\linewidth}p{0.30\linewidth}}
    \toprule
    {\bf Alias} & {\bf Age range} & {\bf Gender identity} & {\bf Country of origin}  \\ 
    \midrule
    JS1  & 60-69  & Male    & Denmark \\ 
    JS2  & 18-29  & Female  & Spain  \\ 
    JS3  & 60-69  & Male    & Denmark  \\ 
    JS4  & 50-59  & Male    & Sweden   \\ 
    JS5  & 30-39  & Male    & South Africa   \\ 
    JS6  & 30-39  & Male    & Italy   \\ 
    JS7  & 30-39  & Female  & Denmark   \\ 
    JS8  & 60-69  & Male    & Denmark   \\ 
    JS9  & 30-39  & Male    & Brazil \\ 
    \bottomrule
    \end{tabular}
    \caption{Collected demographics of job seekers.}
    \label{tab:jsdemoghrphics}
\end{table}

\subsubsection{Job seekers} 

\begin{figure}
   \centering
    \includegraphics[width=0.99\linewidth]{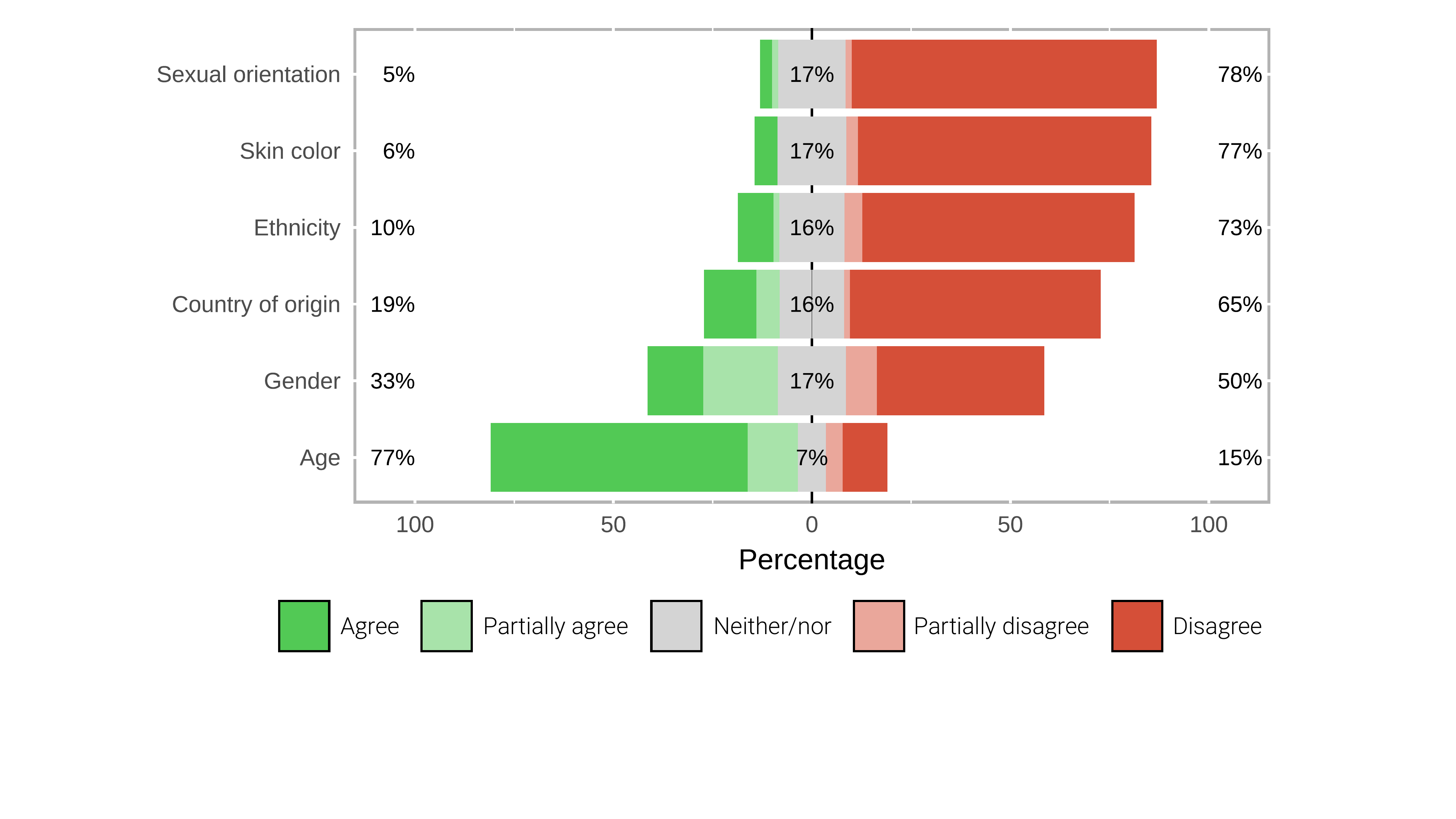} 
   \caption{Experiences with unfairness of the participants in the screener survey used to recruit job seekers ($N$ = 246). All job seekers that had experienced discrimination or unfair treatment while job seeking ($n$ = 74) were asked for the suspected cause: age, gender, country of origin, ethnicity, skin color, and/or sexual orientation.}
   \label{fig:screener}
   \Description[<Job Seeker response>]{<Distribution of the responses of job seekers for being treated unfairly due to protected attributes>}
\end{figure}

In order to recruit job seekers to interview, we shared a screener survey through the recruitment agency's panel of job seekers who have agreed to be contacted by the agency. In addition to demographic questions, we also asked participants whether they had ever experienced discrimination or unfair treatment while job seeking. Out of the 246 responses we received, 30.1\% ($n = 74$) responded affirmatively, 50\% responded negatively and 19.9\% didn't know.
All job seekers that had experienced discrimination or unfair treatment while job seeking ($n$ = 74) were further asked to what extent they thought this mistreatment was due to any of the following sensitive attributes: age, gender, country of origin, ethnicity, skin color, and/or sexual orientation. 
Figure \ref{fig:screener} contains the distribution of their responses for each of these sensitive attributes, and shows that age was by far the most commonly mentioned (suspected) cause of unfair treatment, followed by gender. 
We asked all respondents if they would be interested in being contacted about their personal experiences and opinions on discrimination through an online interview. This resulted in interviews with 9 job seekers, whose demographics are described in \mbox{Table \ref{tab:jsdemoghrphics}}.
%


\begin{table}[]
  \footnotesize
  \centering
  \begin{tabular}{ccp{0.20\linewidth}cp{0.28\linewidth}}
    \toprule
    {\bf Alias} & {\bf Type} & {\bf Interviewee Role} & {\bf \#Employees} & {\bf Industry}\\
    \midrule
    C1  & NGO & HR partner  & 250  & Humanitarian organization\\ 
    C2  & Private  & HR partner & 500  & Gas production\\ 
    C3  & Private  & Customer service manager &  16&Retail\\ 
    C4  & NGO & CEO & 12 & Non-profit organization \\ 
    C5  & Public & HR partner & 700 & Education \\ 
    \bottomrule
    \end{tabular}
    \caption{Collected demographics of companies.}
    \label{tab:companydemographics}
\end{table}

\subsubsection{Companies} 
To recruit companies, we asked the recruitment agency's sales representatives to suggest companies that could potentially be interested in sharing their opinions on and stances towards fairness and bias mitigation. Our goal was to interview both public and private sector companies with a variety of sizes from SMEs to large organizations. However, it proved very difficult to get companies to participate as most of them did not have time or were simply not interested. In the end, we interviewed 5 companies, which are shown in \mbox{Table \ref{tab:companydemographics}}.

\subsection{Interview design}
Our interview guide was largely the same for all stakeholder groups with some minor differences as described below. At the start of each interview, we first explained that their answers would be anonymized and we asked them for permission to record the interviews. Both physical and online interviews were recorded using Microsoft Teams. We also explained that the transcribed interviews would not be shared outside the research teams and that the recordings would be destroyed after 3 months.  We then asked each participant for their consent to participate in the interview and record it. Interviews lasted between 30 to 45 minutes and consisted of the following elements. 

\subsubsection{Introduction} Participants introduced themselves, their roles and who their customers are (if employed at the recruitment agency), and whether they were actively seeking for a job (for job seekers).

\subsubsection{Fairness definition} 
In order not to bias participants towards a specific definition of fairness, we first asked them what fairness meant to them in the context of hiring. Afterwards, we shared our use case-specific definition of fairness (inspired by \citet{holstein2019fairness}): {\em ``Fairness considers the cases where recruiters or the platforms that they use (search engine, recommender system) perform differently for different groups (for example, different age groups, different genders, or people belonging to different racial groups) in ways that may be considered undesirable.''}

\subsubsection{Reflection on hypothetical scenarios} 
To move their ideas and opinions about fairness into the practical realm, we provided each participant with at least two hypothetical scenarios and asked them to reflect on these cases. 
We formulated two different types of scenarios: one related to job seekers and the other related to companies:

\begin{description}

    \item[Scenario 1: Job seeker-focused.] Imagine you are looking for an IT manager\footnote{We selected IT as our default industry as this is an industry where women tend to be strongly underrepresented \cite{heilman2024women}.} for a highly-paid management role and that the list of CVs returned by the search engine or recommender system only contains men in the top results. To find qualified female or non-binary candidates, you find that you need to scroll down all the way to the bottom of the results list. What are your thoughts about this specific example?
    
    \item[Scenario 2: Company-focused.] Imagine there are two similar job postings: one from a well-known and large organization, and the other one from a small, unknown start-up. You realize that the list of recommended or retrieved candidates for the big company includes more relevant candidates. What are your thoughts about this specific example? 
    
\end{description}

Job seekers were provided with two variants of scenario 1, but with different sensitive attributes. At least one of the sensitive attributes came from their indicated experiences in the screener survey.
Recruiters, recruitment agency employees and company representatives were all provided with both scenario types.

\subsubsection{Lived experiences with (un)fairness}
In this part, participants were asked about their lived experiences with (un)fairness in hiring from their perspective. A stakeholder's lived experiences of fairness might differ from theoretical fairness definitions, allowed for the development of algorithms that better align functionality with stakeholder expectations \cite{Smith2024Faact}.
For instance, we asked job seekers to share their personal job seeking experiences, and we asked recruitment agency employees to share their observations based on incoming customer requests and their normal work tasks at the recruitment agency.

\subsubsection{Evaluation brainstorm}
In the final main part of the interviews, we asked each participant to brainstorm on how unfair lists of recommended candidates could be evaluated, thereby helping us co-design evaluation setups and, in some cases, metrics. 
We provided each participant with a hypothetical scenario where recruiters are shown a list of 20 qualified candidates suggested automatically by a recommender system. We then asked participants to brainstorm about how these recruiters could and should assess the fairness of these lists, given that they would have access to any information about the shortlisted candidates. We were specifically interested in what kind of signals and data they needed to assess the fairness of this list or compare it to another one.
Some participants followed their brainstorm path towards a specific metric that could be used to measure their notion of unfairness, while others discussed the type of data that would be necessary for evaluation.

\subsubsection{Closing}
We ended the interviews by asking the participants whether they wanted to add anything or had any questions for us.

\subsection{Data analysis}
We used Goodtape.io to transcribe the recorded interviews as it is a third-party transcription app that the recruitment agency has a Data Protection Agreement with. We used NVivo to perform thematic analysis on the transcribed interviews. After the first author created their code book, all authors discussed the codes and agreed on the final version of the code book about the emerging themes. We present the results of the final coding process in the next section.

\section{Results}
\label{sec:results }


\subsection{Fairness definitions}
\subsubsection{Recruiters / Other recruitment agency employees}
%
Most interviewees' understanding of fairness focused on the importance of qualifications for job seekers: they overwhelmingly agreed that it is essential for potential candidates to possess the qualifications described in the job ad. 
Many employees then stated that, after ensuring candidates' qualifications, they should be treated equally regardless of any other candidate characteristics. For instance, R3 stated that {\em ``fairness is no matter where you're from or what your age or your gender is, you can be contacted with a job, if you have the qualifications''}, a sentiment echoed by R4, R6, E7, E9, E12, and E13. 

Both coming from a sales perspective, E5 expressed that {\em personality traits} should also be considered as qualifications, while E11 felt that {\em cultural fit} should also count as a qualification, something that \citet{rasanen2024personality} have argued is a mistake. 
Other job seeker-oriented definitions were less common. For instance, R2 mentioned that jobs  should be exposed to all job seekers, while E15 felt that everyone should have an equal chance---or `equal opportunity' in the case of R9---of applying for and being hired for a job, without being able to articulate the practical implications of this.
One definition that differed from the rest was given by R10, who stated that {\em ``fairness is also not wasting candidates time''} by not sending them job suggestions for which they would not be invited anyway---something job seekers also commonly mentioned (cf.\ Section \ref{sec:jsfairnessdefinition}).
This is related to the point made by E6 from the {\em Job Advertisement} department, who emphasized the importance of providing correct information in the job advertisement to avoid misleading job seekers into (not) applying for the position. 

Two of the sales managers focused on fairness from the companies' perspective. Both E3 and E5 felt it was important to prevent biases in terms of company size or popularity: {\em ``No matter which size of the company you are, you should be able to reach out the same qualified candidates. The work is equally important.''} (E3). 
%

Finally, only one participant, R6, considered multiple stakeholders in their views on fairness: {\em ``If I have to be fair to the candidates I shouldn't be looking at gender etc. There's also fairness to the companies who are buying the product [from the recruitment agency]''}.

\subsubsection{Job Seekers}
\label{sec:jsfairnessdefinition}
Many of the job seekers' perspectives on fairness echoed those made by the recruiters and other employees. 
JS7 and JS9 echoed the opinion that recruiters and companies should only focus on matching qualifications and not on other characteristics such as sensitive attributes (age, gender identity, country of origin, etc.) or personality traits. This was echoed by JS2: {\em ``Fairness is synonym of impartiality or equal treatment towards all people that apply to a specific job position without any discrimination''}.
Similar to E6, JS8 stressed that the job advertisement should be as clear as possible for job seekers to assess whether the position is relevant for them and to avoid wasting their time.
Similar to R2, JS6 also mentioned the importance of exposing the jobs to as many job seekers possible.
A novel perspective on fairness from multiple job seekers (JS1, JS3 and JS6) focused on the importance of getting feedback from the companies: 
{\em ``So fairness is as long as you're getting a fair treatment that they respect you, and also you get a decent reject. So you know well you didn't get the job because this and this. That's also that you're not evaluated by your age and also gender and so forth.'' } (JS1).

\subsubsection{Companies}
Definitions from our participants from different companies had commonalities between definitions given by job seekers and recruitment agency employees. Similar to R2 and JS6, C1 mentioned that fairness means exposing the job to as many job seekers as possible.  C2 and C5, similar to E15 and R9, mentioned job seekers having equal chance to be interviewed or hired: {\em ``Fairness from my perspective is that everybody has an equal chance to get to an interview or get the job in the end. We must not have bias towards names, gender, nationality, education, although in some roles we need a specific education, of course''} (C2). C3, like E11 described cultural fit as fairness, by focusing on ``what is best for the team''.

\subsection{Reflection on hypothetical scenarios}

\subsubsection{Job seeker-focused scenario}
Participants varied in their reactions to the hypothetical job seeker-focused scenario. While several participants found the hypothetical candidate lists to be problematic, some attempted to justify how such lists could happen. 
JS5, JS7, JS8, C4, C5 and R9 all found the skewed lists problematic, as evidenced by R9's statement: {\em ``That's very stereotypical, that we think a good leader equals masculine qualities. The engine [search engine/recommendations] shouldn't be that way''}. 
%
E1, R4, R7, R8 and JS6 attempted to justify why these lists appeared skewed. E1, for instance, thought the ranking was purely the result of the listed candidates being more qualified and that sensitive attributes did not play a role in this.
In contrast, R4 and R8 believed the rank of the returned candidates is based only on their search criteria.
None of the recruiters, however, reflected on whether they would be able to detect whether a candidate list was {\em too} skewed on their own.

JS6 felt that such skewed candidate lists are only acceptable if the company has a valid reason for candidates having to belong to a specific group (e.g., age or gender). 
This matched one of the emergent themes of specific {\em types of job} being dominated by candidates matching a specific attribute, with gender being the most common example. C2, C4, JS2, JS3, JS8, E5, E6 and E9 all focused on this aspect. For example, for the highly-paid management IT position C4 stated: {\em ``I guess that it doesn't come as a surprise, at least in the sense that we are a very gender-divided labor market and we have a lot of specific working areas that are linked to either one or the other gender. IT being a male dominated field''}. Other example professions that were seen as male-dominated included auto mechanic (E6), blacksmith (JS3), and special engineer (C2), while nursing was seen as an example of a female-dominated profession (JS7). 

JS2 reflected on a similar imbalance related to age: {\em ``If you are looking for a senior executive, then of course, it would be very weird to me that no one is 50+, but it would make total sense that there are no newly graduates, of course. So I think it really depends on the position that you want to to fill.''} 
E5 and JS7 emphasized the importance of sensitive attribute distribution being proportional to the population of the qualified candidates available in the CV database: {\em ``If, let's assume we have in our CV database, there are 25 qualified candidates, and five out of the 25 are women, then it is normal to have[=contact] 1 out 5 candidates as women''}. 

Finally, C3, JS3 and JS4 focused on diversity. While JS3 and JS4 felt the list should be diverse with respect to age, C3 stressed the importance of interviewing an overall diverse set of candidates: {\em ``In the interview process you want a diverse set of candidates to consider for the position. Because I think it sometimes gets better in the working environment if we are different; we give something each of us''}.

\subsubsection{Company-focused scenario}

Fairness in algorithmic hiring from the company perspective brings its own challenges, such as company size influencing the way they are treated and conceived of by recruiters and job seekers.
Both E1 and R1 explained how, even for the same type of position, the size of the company can influence what is required from the ideal candidate, with smaller companies needing more all-round and less specialized employees for the same position. 
R1 stated job titles could vary for the same position: {\em ``When you're looking at the job market, it would not typically be the same job title in a small firm than in a big firm''}. 
%
E3 believed the choice of job title could also affect the performance of the recruiter: {\em ``They will not be aware of it, but they may be more serious about the big brand job. We want everybody to work at big company and they're a good customer. So they do a better job. Then with a small company, maybe they think they're only gonna buy a few times from us [= the recruitment agency], so why do I have to put in a lot of work?''}. 
This suggests that recruiters could be influenced by their familiarity with the company, which could affect their performance and result in unfair treatment of smaller companies.  


\subsection{Lived experiences with (un)fairness}
\label{sec:livesExperiences}
In this section, we summarize some of the common examples of lived experiences with (un)fairness from our stakeholders. Although not used directly to select fairness metrics, we include them to show the variety of biases that can result in unfair situations in hiring.
Recruiters identified several factors that can create biases,  including recruiter experience, the tools they use, time limits imposed on their recruiting tasks, and the way job ads are worded among others. One example, related to recruiter experience and the tools they use, involved a new recruiter not knowing which search keywords to use, potentially resulting in relevant candidates being left uncontacted. Another example was that when recruiters face a higher-than-usual workload, they may spend less time on each job, which again can lead to relevant candidates being overlooked.

Recruitment agency employees with direct customer contact provided several examples of customers expressed preferences for specific subgroups of job seekers related to sensitive attributes, such as age and gender. A common example was customers stating they did not want newly graduated candidates or candidates close to retirement age. Another example was some companies preferring no female candidates with small kids or at the age of potentially going on maternity leave, suggesting issues with intersectional fairness \cite{Vethman2025Faact}. Job seekers mostly gave examples where they felt they were being treated unfairly due to age, gender identity, ethnicity, or family life stage (such as pregnancy or small kids). Finally, company representatives expressed problems with so-called {\em similarity attraction bias} \cite{rivera2012hiring}, where they made the mistake of hiring someone similar to the other team members and that would fit into the company culture.

\subsection{Evaluating \& measuring fairness}
In the final part of the interviews, we asked participants to explain how they would evaluate the fairness of a hypothetical list of 20 recommended candidates, thereby helping us co-design evaluation setups and, in some cases, metrics. This section describes the different signals and properties the participants mentioned.

\subsubsection{Qualifications} Almost all of our participants emphasized the importance of selecting qualified candidates. While a handful of participants (C3, JS3, R6, E15) mentioned that matching the job's required qualifications was the {\em only} important signal, a majority of the other participants proposed using {\bf being qualified} as a necessary precondition to even start looking at other signals relevant for measuring fairness. 

\subsubsection{Size of the candidate pool}
An oft-mentioned factor in assessing the fairness of a candidate list was the {\bf size of the candidate pool}. C1 and JS9 explained that smaller candidate pools make it harder to achieve fairness. For example, JS9 said: {\em ``The size of the candidate pool is another important thing to look at, because for some type of jobs, maybe there are hundreds of them [= qualified candidates], and if we are planning to just interview 10, then randomizing is easier. But if it was a job that only five people [qualified for], then you would interview everybody, no matter what''}, implying that fairness could not be guaranteed in such a situation, a sentiment echoed by R4.
%
Related to this, E7 explained how overly large pools could create other types of unfair situations as time limitations could make it difficult to assess all the candidates in the pool: {\em ``There was this job as a Dynamic.NET Vision Specialist, and we had nine of them at the moment in the database. But of course, it's much more different if you're looking for a receptionist. Then I think you can come to the point where you haven't scrolled them all through''}. 
According to JS6, business constraints could also cause such unfair situations: {\em ``If you show a little group out of this big pool of people, you risk to be unfair. Your risk is to be highly unfair because it's a small number. And then you say, well, then I increase the number. But then the recruiters and the company might say, I cannot invest, you know, one month just to screen 100 people just because I wanted the screening kind of there's fairness inside. So I understand like companies need to do business''}.

\subsubsection{Distribution of qualified candidates with respect to sensitive attributes}
Another common theme by different stakeholders was the comparison of the set of shortlisted candidates to the set of all available job seekers with regard to the {\bf distribution of a sensitive attribute}. C2, C5, E10, R8, JS2, JS7 and JS9 all provided different examples, mostly focusing on jobs with skewed gender distributions. For example, R8 said: {\em ``If we had the data for what are the pool of the candidates available for this position compared to the list of 20 people I had chosen. And this list should be representative of the [larger] pool. Because if the pool is like 50/50, and my list [has] 18 women, then there is a problem''}. 
For example, JS7 raised the issue with nursing jobs being a female-dominated profession: {\em ``Based on the fact that 50\% of nurses are not men, if you took 50\% men and 50\% women, there would be an over-representation of men relative to how many there are [in the pool of nurses]. So I think it's fair that there is an over-representation if there is an over-representation in the group you are looking for''}. 
E10 focused on ethnicity instead of gender: {\em ``I think maybe based on the CVs and based on that you can maybe see in a way what have been sent out by minorities and all, you know how many you got from which minority group''}.

\subsubsection{Rank-awareness}
Recruiters that have experience with searching for candidates using the recruitment agency's CV search engine or assessing the list of recommended candidates stressed the importance of the candidates' {\bf rank} in assessing their relevance. In line with established models of browsing \cite{carterette2011system}, they suggest that as recruiters go down the ranked results lists they risk growing less interested with each additional candidate to inspect. 
R8 explains that recruiters expect candidates shown near the top of the list to be more relevant. 
E7 is one of the few participants who is aware that some potentially relevant candidates may never be shortlisted and  offered the position simply because the algorithm places them lower in the list: {\em ``I start from the top, and find 30 qualified candidates. Then on page two, there could be as many qualified candidates, but they won't get the offer because they're not in the top''}. R6 admits that, in line with \cite{carterette2011system}, this could be due to fatigue: {\em ``If I scroll through CVs, maybe around CV number 20, I would get distracted.''} 

\subsubsection{Diversity}
Nearly half of the job seekers (JS1, JS3, JS5 and JS9) mentioned {\bf diversity} in terms of one or more sensitive attributes as an important aspect. For instance, JS1 said: {\em ``You should work out a diverse list. What is diversity? That's age, that's gender, and I think that's the country of origin''}. JS5 argued that diversity depends on the type of job, while C1 argues it is also dependent on the size of the candidate pool: {\em ``If we just took the first 30 [job seekers] that could actually do the job. But if the next 20 is the same qualified as the other on paper, I would like to mix it a little bit if there were other types of candidates there''}.

According to JS3 and JS5, what this diversity should look like depends not only on the candidate pool but also on the diversity of the team that the preferred candidate would join at the company offering the position---or so-called {\em cultural add}---and the company's broader mission: 
{\em ``Some companies want to give more chances to young people, or to senior people that left the job market that are trying to come back now. It depends on this kind of mission-oriented approach that a company might want to take as well. Some do have social missions and they want to execute that through hiring process.''} (JS9). 
C4 provided an important insight on {\em cultural add} vs {\em cultural fit} from the companies' perspective: {\em ``I love when there's a cultural fit. It's so easy. Cultural add is more challenging. But I also believe that with the right type of workplace, it can really create some good things''}.

One participant stressed out that fairness can be measured by looking at the coverage of the job seekers: {\em ``We can measure simply do we get more applicants overall? And what can we say about these applicants?''} (E4). Coverage is one type of aggregate diversity \cite{Mansoury2020FairMatch}.


\subsubsection{Equal quotas}
Only JS4 among our participants fully supported the idea of {\bf equal quotas} and C5 noted it may be necessary for some fields: {\em ``I think in some areas, it is the only way to get some employees or management to get them aware of the bias that there is in the organization and to get them to rethink their competences. I think in some cases it is necessary''}. 
In contrast, JS6 and JS7 explained how aiming for strict equal quotas can create other types of unfair situations. For instance, JS6 said: {\em ``If you strictly want an attribute to represent strict quota, that may exclude some other attributes that are more important for your job, and then you lose qualified people''}.

\subsubsection{Broadening fairness evaluation}
A final notion worth mentioning is that it could be important to consider not only the shortlisted and contacted candidates when measuring fairness, but also {\bf broaden} it to the set of candidates that actually applied for the job (R9) as well as the set of job seekers that were not shortlisted (JS7). 

\subsection{Mapping findings to fairness metrics}
\label{sec:mappingtometrics}

\subsubsection{Individual fairness}
When we asked participants to share their own definition of fairness and when we asked them to reflect on the hypothetical scenarios, our participants' notions of fairness overwhelmingly corresponded to {\em individual fairness} \cite{castelnovo2022clarification,rampisela2023fairness}. Individual fairness corresponds to the principle that similar individuals should be given similar decisions, and focuses on the comparison of single individuals instead of groups of people sharing some characteristics \cite{castelnovo2022clarification}. 
We observed two interesting patterns in their definitions. First, their initial notions of individual fairness was strongly tied to qualifications: as long as candidates were qualified, many participants seemed oblivious to other fairness issues. Second, a majority of their fairness definitions were centered around the job seekers, with some definitions focused on the companies as a stakeholder. Perhaps due to the special consumer role that recruiters play in our algorithmic hiring scenario, no definitions emerged centered around the recruiters.

\emph{\textbf{Job seeker-focused fairness}}
Our definition of individual fairness for job seekers follows the following principle: {\em similar individual job seekers should be subject to similar decisions}, either by human recruiters or the decision-making tools they use, such as recommender systems or CV search engines. We combine our different stakeholders' definitions for job seeker fairness into the following definition: {\em ``If there are $\mathcal{N}$ qualified candidates among the set of available job seekers for a job, then they should all have the same probability of being {\em exposed to the recruiters}, being {\em contacted for that job}, or {\em applying for the job}, regardless of their sensitive attributes.''}

There are two important open questions this definition requires an answer to. The first is on what basis should individual job seekers be considered similar? In other words: in addition to qualifications, which sensitive attributes---age, gender, sexual orientation, disability, ethnicity, race, skin color, or country of origin---should be incorporated in this, and how do we deal with proxies for these attributes? The second question is what makes a job seeker qualified for a given job (and who decides this)?  

\emph{\textbf{Company-focused fairness}}
Our definition of individual fairness for companies follows the following  principle: {\em similar job postings should be subject to similar decisions}. In other words, similar sets of job seekers should be shortlisted by human recruiters or their decision-making tools for similar jobs. We combine the definitions for company fairness given by the different stakeholders into the following definition: {\em ``If there are similar job positions coming from  $\mathcal{M}$ different companies, then all $\mathcal{M}$ companies have the same probability to get qualified candidates [= receiving the shortlist provided by the recruiters] regardless of company size or popularity.''}

An important open question here is how to identify {\em which} pairs of job postings are similar enough for them (and thereby their companies) to be treated the same and given the same chance to be recommended to qualified candidates? 
 
\subsubsection{Group fairness}
Interestingly, when we moved on to interview phase where  participants were asked to brainstorm about what they would need to evaluate the fairness of a list of qualified candidates, we noticed their notions of fairness shifted more towards {\em group fairness}. Group fairness is centered around the idea that some groups of people potentially suffer biases and unfair decisions, and therefore attempts equality of treatment for groups instead of individuals \cite{castelnovo2022clarification}.
The discussion around hypothetical scenarios, and brainstorming session for measuring fairness emerged {\em group fairness} metrics, which in general aims to reach equality of treatment for groups (for example groups based on different values of gender identity for job seekers, and groups based on different values of company size for companies) instead of individuals. 

\subsubsection{Mapping signals to categories of fairness metrics}
In this section, we attempt to direct the results from the evaluation brainstorm phase directly to (families of) fairness metrics.

\emph{\textbf{Distribution of qualified candidates}}
Many of our participants felt it essential to compare the distribution of shortlisted candidates to the available pool of job seekers with regard to sensitive attributes. This notion maps directly to {\em Conditional Demographic (Dis)Parity} (CDP)\cite{Watcher2021fairness,castelnovo2022clarification}, which aims to measure the equal acceptance rate across groups in any strata. For instance, to achieve fairness in our candidate recommendation scenario, we could choose gender identity as the protected attribute. This means that, even though we still want to recruit the most skilled candidates, it would require the decision-making (e.g., recommending, shortlisting, contacting, hiring) to be independent of the candidate's gender identity, but conditional on being qualified. 
%
The majority of our interviewed stakeholders ---from job seekers and company representatives to recruiters and other agency representatives---
focused on the importance of qualifications and also the distribution of the candidates in the pool of candidates, which all are properties of CDP.  
We should note that using ``qualifications'' as a condition to measure fairness might itself contribute to unfairness. For example, due to historical factors and societies that are historically male-dominated, the skills, work experience and educational background of female candidates may differ from those of male candidates, leading to potentially unfair outcomes when focused (solely) on qualifications \cite{castelnovo2022clarification}.

\emph{\textbf{Rank-awareness}}
Rank-awareness is also mentioned as a signal to use to measure fairness, especially for the tools that the recruitment agency recruiters are using to identify relevant candidates: ranked candidate lists from CV search engines and candidate recommender systems. 
A metric that could be used to quantify the bias in such ranked outputs 
is {\em skew} \cite{geyik2019fairness}, which computes the extent to which a ranked list of candidates for a search or recommendation task differs over a sensitive attribute value with respect to the desired proportion of that attribute value in the pool of available candidates. Other rank-aware fairness metrics that consider comparing distributions have been also proposed and could be considered here \cite{yang2017measuring, raj2022measuring}.

\emph{\textbf{Diversity}}
It is not surprising that multiple participants focused on diversity as a way to evaluate fairness. As \citet{zhaofairnessdiversity2025} discussed, differential treatment towards diverse groups can cause bias, and the notion of (group) fairness requires us to provide similar treatments to these diverse groups. Metric-wise, this maps to {\em category-based diversity} metrics \cite{zhaofairnessdiversity2025}, which consider sensitive attributes as categories for measuring diversity, such as gender. 

\emph{\textbf{Coverage}}
The percentage of items that can be recommended by the system is also mentioned by our participants as something to measure in relation to fairness. The coverage metric \cite{Mansoury2020FairMatch} is a type of {\em qualification fairness} metric \cite{rampisela2023fairness}, from the family of individual fairness metrics. In our algorithmic hiring scenario, it could be used to compute the percentage of available job seekers that are recommended, returned by the search engine, contacted by recruiters and so on. To consider rank-awareness, the {\em Gini index} \cite{rampisela2023fairness} could also be used, which measures to what degree the distribution of items in a ranked list deviates from an equal/uniform distribution.  

\emph{\textbf{Company parity}}
While a majority of our participants focused on fairness for the job seeker, some participants also highlighted the importance of fairness from the company perspective. To measure fairness from the company perspective, we could adopt 
fairness metrics that aim to measure disparate treatment of the company groups with company size as our protected attribute. For instance, the {\em deviation from consumer fairness parity (DCF)} can be used to compare the average number of contacted candidates for different company groups, which is a form of non-parity fairness \cite{yao2017beyond}.

\section{Discussion \& Conclusions}
\label{sec:discussion}

In this study, we conducted 40 semi-structured interviews with different stakeholders of an algorithmic hiring system focused on candidate recommendation: job seekers, company representatives, recruiters and other employees of a recruitment agency. After exploring their lived experiences, definitions of fairness, and how to evaluate it, we attempted to map our findings based on stakeholders' needs to fairness definitions and potentially fairness metrics. To the best of our knowledge, we are the first to perform such a mapping based on the qualitative inputs of multiple stakeholder groups in the domain of algorithmic hiring.

Our interviews show that, across the different stakeholders in the hiring domain, preconceived notions of fairness closely match the principles of individual fairness with a strong emphasis on qualifications and a trust in the systems generating these lists without much reflection on any biases they may contain. Nevertheless, our stakeholders were able to share many lived experiences of unfairness with us and factors that could have caused them.
%
%
However, after being asked to reflect on hypothetical lists of candidates and brainstorm about how best to evaluate them for fairness, we saw their notions of fairness shift more towards the idea group fairness, especially in terms of protected attributes such as gender and age for job seekers, and size for companies. 
Our analysis enabled us to map these findings to relevant fairness metrics, with Conditional Demographic (Dis)Parity emerging as one of the key measures. Other important properties included rank-awareness and diversity. 
%


It is important to note that the work presented in this paper was prompted by the EU AI Act. While other fairness-related regulation---such as the NY FAIR Business Practices Act---may differ from the EU AI Act in scope, we believe that any regulation with the aim to ensure fairness for consumers or other stakeholders should consider a multi-stakeholder approach to identifying the right fairness metrics. Our paper can be a guide on how to successfully conduct such an multi-stakeholder approach.

\emph{\textbf{Limitations}}
Our study has a number of limitations. First, we note that the sample of recruited job seekers could have been more diverse in terms of in gender identity and racial/ethnicity identity, and the results presented in the rest of the paper should be interpreted accordingly. Similarly, we also note the lack of diversity in the companies that were interviewed as a limitation of this research study, and our results should be interpreted with this in mind. 
Second, although we map perspectives of stakeholders to fairness metrics, we have not yet conducted an empirical analysis of job recommendation outcomes using these metrics. We leave this for future work.
Finally, our study specifically focused on fairness in algorithmic hiring recommender systems and our participant sample is mainly from stakeholders in Denmark who have interacted with a particular recruiting agency. Hence, our study might not take into account constraints in other domains, which could impact the generalizability of our findings beyond HR domain and our findings are culture and country specific. 

\emph{\textbf{Future work}}
While there are many avenues for future work, we describe a concrete one here: using our knowledge of the different stakeholder perspectives on fairness and the corresponding sets of fairness metrics to aid the recruitment agency in designing an algorithmic auditing framework to perform (periodic) audits of their algorithmic hiring systems, as mandated by EU AI Act in such a high-risk domain.
%
We will develop an auditing framework for the recruitment agency 
to document the intended use, training data, data sources, algorithmic performance, and optimization goals of recruitment agencies' algorithmic hiring workflows \cite{raji2020closing}. 
The unique advantage of the human-augmented algorithmic hiring setup at this recruitment agency is that we can use our final set of fairness metrics to identify fairness issues both in our candidate recommender system and in the recruiters' manual matching practice. This could lead to interesting insights into how the interplay between human and algorithm affects fairness.

\emph{\textbf{Reproducibility}} Interview guides and codebook based on thematic coding in this paper is available online at \url{https://github.com/mesutkaya/recsys2025}.

\begin{acks}
This work was supported by FairMatch project (Innovation Fund Denmark grant number 3195-00003B). The work of Toine Bogers was also supported by the Pioneer Centre for AI, DNRF grant number P1. 
We would like to thank Jobindex A/S and its employees for their valuable participation in our qualitative study and for providing communication channels that enabled us to interview both company representatives and job seekers. We are also deeply grateful to the job seekers and company representatives who generously shared their time and insights as part of this research.
    
\end{acks}

\bibliographystyle{ACM-Reference-Format}
\bibliography{recsys2025-multistakeholder-fairness}

\end{document}